\documentstyle[aps]{revtex}

\begin{document}
\title{Pairing in the Three-Band Hubbard Model of the Cu-O Plane}
\author{Michele Cini, Gianluca Stefanucci and Adalberto Balzarotti}
\address{INFM, Dipartimento di Fisica,Universit\`{a} di Roma Tor Vergata, Via\\
della Ricerca Scientifica 1- 00133 Roma, Italy}
\date{\today}
\maketitle

\begin{abstract}
By a canonical transformation of the three-band Hubbard model, we introduce
an effective Hamiltonian for the propagation of two holes doped into the
ground state of the Cu-O plane. When the pair belongs to the $^{1}B_{2}$ or $%
^{1}A_{2}$ Irreducible Representations of the C$_{4v}$ Group, the bare holes
do not interact by the on-site repulsion; the effective interaction between
the dressed holes is obtained analytically in terms of renormalized matrix
elements, and generalizes earlier findings from cluster calculations. The
Fermi liquid is unstable and numerical estimates with reasonable parameters of
the binding energy of the pair are in the range of tens of meV. Our scheme
naturally lends itself to embody phonon-mediated and other interactions which
cannot occur in the Hubbard model but may give important contributions.
\end{abstract}

\pacs{P.A.C.S
numbers:
74.72-h High-Tc cuprates;
31.20.Tz  Electronic correlation and CI calculations;
74.20.-z
Theory
of
superconductivity }

\noindent

\narrowtext
%\section{Theory}

%\section{Theory}
The discovery of high-$T_{C}$ superconductivity\cite{kn:bm} has stimulated
several serious and ingenious proposals for the pairing mechanism, ranging
from enhanced BCS-type\cite{kn:pietro} and polaron\cite{kn:iad} to a variety
of purely electronic ones\cite{kn:bob,kn:pw,kn:sus}, just to cite a few of
the less exotic proposals. These proposals have usually been regarded as
alternative, but there is a clear need for a theoretical framework where
they can coexist and can be compared on equal footing, since all the serious
mechanisms must convey part of the physics. Our first aim here is to propose
such a scheme. Our approach started with cluster calculations in the
three-band Hubbard model; pioneers in this field\cite{kn:hirsch,kn:bal}
explored the possibility that pairing could result from repulsion; later it
was pointed out\cite{kn:cb1} that W=0 pairs have intriguing properties. W=0
pairs are two-hole eigenstates of the kinetic energy $H_{0}$ that are also
eigenstates of the on-site repulsion term $W$ with eigenvalue 0. The
symmetry of the cluster is an essential ingredient, because only fully
symmetric clusters allow such solutions. The planar lattice structure is
also essential, because no W=0 pairs occur in 3D or in a continuous model.
The cluster calculations\cite{kn:cb2,kn:cb3,kn:cb4} showed that W=0 pairs
are the $^{\prime \prime }bare^{\prime \prime }$ quasiparticles that, when $%
^{\prime \prime }dressed^{\prime \prime }$, become a bound state. That
approach is inherently limited by the small size of solvable clusters, but
allows a very explicit display of paired hole properties, that even show
superconducting flux-quantization\cite{kn:cb5}. The main indirect
interaction through the background particles arises from the second-order
contributions to the two-hole amplitude\cite{kn:cb5}. Our second task here is to
extend the theory to the full plane. The three-band Hubbard model Hamiltonian is
$ H=H_{0}+W$, and the independent hole hamiltonian reads, in the site
representation

\begin{equation}
H_{0}={\sum_{Cu}}\varepsilon _{d}n_{d}+{\sum_{O}}\varepsilon _{p}n_{p}+{\
t\sum_{n.n.}}\left[ c_{p}^{+}c_{d}+h.c.\right]  \label{1}
\end{equation}

where n.n. stands for nearest neighbors. The on-site repulsion Hamiltonian
will be denoted by 
\begin{equation}
W={\sum_{i}}U_{i}n_{i+}n_{i-};  \label{2}
\end{equation}

where $U_{i}=U_{d}$ for a Cu site, $U_{i}=U_{p}$ for an Oxygen. Omitting the
band indices, we shall mean 
\begin{equation}
d[k]=\left\| k_{+},-k_{-}\right\| =c_{k,+}^{+}c_{-k,-}^{+}|vac>  \label{3}
\end{equation}
to be a two-hole determinantal state derived from the k eigenfunctions.

The point symmetry Group of the Cu-O plane is $C_{4v}$, and its character
Table is shown in Table 1. We introduce the determinants $Rd[k]=d[Rk],R\in
C_{4v}$ , and the projected states 
\begin{equation}
\Phi _{\eta }\left[ k\right] =\frac{1}{\sqrt{8}}{\sum_{R\in C_{4v}}}\chi
^{\left( \eta \right) }\left( R\right) Rd[k]  \label{4}
\end{equation}
where $\chi ^{\left( \eta \right) }(R)$ is the character of the operation $R$
in the Irreducible Representation (Irrep) $\eta $. In the non-degenerate
Irreps, the operations that produce opposite $Rk$ have the same character,
and the corresponding projections lead to singlets. Let $R_{i},i=1,..8$
denote the operations of $C_{4v}$ and $k,k^{\prime} $ any two points in the
Brillouin Zone (BZ). Consider any two-body operator $\hat{O}$, which is
symmetric ($R_{i}^{+}\hat{O}R_{i}=\hat{O}$), and the matrix with elements $%
O_{i,j}=<d[k]|R_{i}^{+}\hat{O}R_{j}|d[k^{\prime }]>$, where $k$ and $%
k^{\prime }$ may be taken to be in the same or in different bands. This
matrix is diagonal on the basis of symmetry projected states, with
eigenvalues 
\begin{equation}
O\left( \eta ,k,k^{\prime }\right) ={\sum_{R}}\chi ^{\left( \eta
\right) }\left( R\right) O_{R}\left( k,k^{\prime }\right)   \label{5}
\end{equation}
where 
\begin{equation}
O_{R}\left( k,k^{\prime }\right) =\left\langle d[k]|\hat{O}
|Rd[k^{\prime }]\right\rangle .  \label{6}
\end{equation}
Thus, omitting the $k$, $k^{\prime }$ arguments, we get in particular 
\begin{eqnarray}
O\left( ^{1}A_{2}\right)  &=&O_{E}+O_{C_{2}}+O%
_{C_{4}}+O_{C_{4}^{3}}  \nonumber \\
&&-O_{\sigma _{x}}-O_{\sigma _{y}}-O_{\sigma _{1}^{\prime
}}-O_{\sigma _{2}^{\prime }}  \label{7}
\end{eqnarray}
\begin{eqnarray}
O\left( ^{1}B_{2}\right)  &=&O_{E}+O_{C_{2}}-O%
_{C_{4}}-O_{C_{4}^{3}}  \nonumber \\
&&-O_{\sigma _{x}}-O_{\sigma _{y}}+O_{\sigma _{1}^{\prime
}}+O_{\sigma _{2}^{\prime }}  \label{8}
\end{eqnarray}
If $\hat{O}$ is identified with W, since $W_{E}=W_{C_{2}}=W_{\sigma
_{x}}=W_{\sigma _{y}}$ and $W_{C_{4}}=W_{C_{4^{3}}}=W_{\sigma _{1}^{\prime
}}=W_{\sigma _{2}^{\prime }}$, one finds $W\left( ^{1}A_{2}\right) =W\left(
^{1}B_{2}\right) =0$ . These are W=0 pairs, like those studied previously%
\cite{kn:cb5}. One necessary condition for pairing in clusters is that the
least bound holes form such a pair, and this dictates conditions on the
occupation number. In the full plane, however, W=0 pairs exist at the Fermi
level for any filling.

Suppose the Cu-O plane is in its ground state with Fermi energy $E_{F}$ and
a couple of extra holes are added. In principle, by a canonical
transformation one can obtain an effective Hamiltonian which describes the
propagation of a pair of \underline{dressed} holes, and includes all
many-body effects. Let us see how this arises.

The exact many-body ground state with two added holes may be expanded in
terms of excitations over the vacuum (the non-interacting Fermi {\em sphere}%
) by a configuration interaction: 
\begin{equation}
|\Psi _{0}>={\sum_{m}}a_{m}|m>+{\sum_{\alpha }}b_{\alpha }|\alpha >+{\
\sum_{\beta }}c_{\beta }|\beta >+....  \label{9}
\end{equation}
here m runs over pair states, $\alpha $ over 4-body states ($2$ holes and 1
e-h pair), $\beta $ over 6-body ones ($2$ holes and 2 e-h pairs), and so on.
To set up the Schr\"{o}dinger equation, we consider the effects of the
operators on the terms of $|\Psi _{0}>$. We write:

\begin{equation}
H_{0}|m>=E_{m}|m>,H_{0}|\alpha >=E_{\alpha }|\alpha >,...  \label{10}
\end{equation}

and since W can create or destroy up to 2 e-h pairs,

\begin{eqnarray}
W|m>={\sum_{m^{\prime }}}V_{m^{\prime },m}|m^{\prime }>+{\sum_{\alpha }}%
|\alpha >W_{\alpha ,m}  \nonumber \\
+{\ \sum_{\beta }}|\beta >W_{\beta ,m}.  \label{11}
\end{eqnarray}
$V_{m^{\prime },m}$ vanishes for W=0 pairs in our model; however we keep it
for generality, since it allows to introduce the effect of phonons and any
other indirect interaction that we are not considering. For clarity let us
first write the equations that include explicitly up to 6-body states; then
we have 
\begin{eqnarray}
W|\alpha &>&={\sum_{m}}|m>W_{m,\alpha }+{\sum_{\alpha ^{\prime }}}|\alpha
^{\prime }>W_{\alpha ^{\prime },\alpha }  \nonumber \\
+{\sum_{\beta }}|\beta &>&W_{\beta ,\alpha }  \label{12}
\end{eqnarray}

where scattering between 4-body states is allowed by the second term, and

\begin{eqnarray}
W|\beta >={\sum_{m^{\prime }}}\left| m^{\prime }\right\rangle W_{m^{\prime
},\beta }+{\sum_{\alpha }}\left| \alpha \right\rangle W_{\alpha ,\beta }
\nonumber \\
+{\sum_{\beta^{\prime } }}\left|\beta^{\prime } \right\rangle W_{
\beta^{\prime } ,\beta }
\label{13}
\end{eqnarray}
In principle, the $W_{\beta^{\prime } ,\beta }$ term can be
eliminated by taking linear combinations of the complete set of $\beta $~ states:
when this is done, we get a self-energy correction to $E_{\beta }$ and a
renormalization of the vertices, without altering the structure of the
equations. The Schr\"{o}dinger equation yields equations for the coefficients a,b
and c 
\begin{eqnarray}
\left( E_{m}-E_{0}\right) a_{m}  \nonumber \\
+{\sum_{m^{\prime }}}a_{m^{\prime }}V_{m,m^{\prime }}+{\sum_{\alpha }}%
b_{\alpha }W_{m,\alpha }+{\sum_{\beta }}c_{\beta }W_{m,\beta } &=&0
\label{14}
\end{eqnarray}

\begin{eqnarray}
\left( E_{\alpha }-E_{0}\right) b_{\alpha }  \nonumber \\
+{\sum_{m^{\prime }}}a_{m^{\prime }}W_{\alpha .,m^{\prime }}+{\sum_{\alpha
^{\prime }}}b_{\alpha ^{\prime }}W_{\alpha ,\alpha ^{\prime }}+{\sum_{\beta }%
}c_{\beta }W_{\alpha ,\beta } &=&0  \label{15}
\end{eqnarray}

\begin{equation}
\left( E_{\beta }-E_{0}\right) c_{\beta }+{\sum_{m^{\prime }}}a_{m^{\prime
}}W_{\beta ,.m^{\prime }}+{\sum_{\alpha ^{\prime }}}b_{\alpha ^{\prime
}}W_{\beta ,\alpha ^{\prime }}=0  \label{16}
\end{equation}

where E$_{0}$ is the ground state energy. Then, we exactly decouple the
6-body states by solving the equation for $c_{\beta }$ and substituting into
the previous equations, getting:

\begin{eqnarray}
\left( E_{m}-E_{0}\right) a_{m}+{\sum_{m^{\prime }}}a_{m^{\prime }}\left[
V_{m,m^{\prime }}+{\sum_{\beta }}\frac{W_{m,\beta }W_{\beta ,m^{\prime }}}{
E_{0}-E_{\beta }}\right]  \nonumber \\
+{\sum_{\alpha }}b_{\alpha }\left[ W_{m,\alpha }+{\sum_{\beta }}\frac{
W_{m,\beta }W_{\beta ,\alpha }}{E_{0}-E_{\beta }}\right] =0  \label{17}
\end{eqnarray}

\begin{eqnarray}
\left( E_{\alpha }-E_{0}\right) b_{\alpha }+{\sum_{m^{\prime }}}%
a_{m^{\prime }}W_{\alpha .,m^{\prime }}  \nonumber \\
+{\sum_{\alpha ^{\prime }}}b_{\alpha ^{\prime }}\left[ W_{\alpha ,\alpha
^{\prime }}+{\sum_{\beta }}\frac{W_{\beta ,\alpha ^{\prime }}W_{\alpha
,\beta }}{E_{0}-E_{\beta }}\right] &=&0  \label{18}
\end{eqnarray}
Thus we see that the r\^{o}le of 6-body states is just to renormalize the
interaction between 2-body and 4-body ones, and for the rest they may be
forgotten about. If $E_{0}$ is outside the continuum of excitations, as we
shall show below, the corrections are finite, and experience with clusters
suggests that they are small. Had we included 8-body excitations, we could
have eliminated them by solving the system for their coefficients and
substituting, thus reducing to the above problem with further
renormalizations. In principle, the method applies to all the higher order
interactions, and we can recast our problem as if only 2 and 4-body states
existed. Again, the $W_{\alpha^{\prime } ,\alpha }$ term can be
eliminated by taking linear combinations of the $\alpha $~ states: when this
is done, we get a self-energy correction to $E_{\alpha }$ and a
renormalization of the $W_{m,\alpha }$ vertices. The equations become

\begin{equation}
\left( E_{m}-E_{0}\right) a_{m}+{\sum_{m^{\prime }}}a_{m^{\prime
}}V_{m,m^{\prime }}+{\sum_{\alpha }}b_{\alpha }W_{m,\alpha }=0  \label{19}
\end{equation}

\begin{equation}
\left( E_{\alpha }-E_{0}\right) b_{\alpha }+{\sum_{m^{\prime }}}a_{m^{\prime
}}W_{\alpha ,m^{\prime }}=0  \label{20}
\end{equation}
Solving for $b_{\alpha }$ and substituting in the first equation we exactly
decouple the 4-body states as well. The eigenvalue problem is now 
\begin{equation}
\left( E_{0}-E_{m}\right) a_{m}=\sum_{m^{\prime}} a_{m^{\prime }}\left\{
V_{m,m^{\prime }}+\left\langle m|S[E_{0}]|m^{\prime }\right\rangle \right\} , 
\label{21}
\end{equation}

where

\begin{equation}
\left\langle m|S\left[ E_{0}\right] |m^{\prime }\right\rangle ={\sum_{\alpha
}}\frac{<m|W|\alpha ><\alpha |W|m^{\prime }>}{E_{0}-E_{\alpha }}.  \label{22}
\end{equation}

This is of the form of a Schr\"{o}dinger equation with eigenvalue $E_{0\text{
}}$for pairs with an effective interaction $V+S.$ Then we interpret $a_{m}$%
as the wave function of the dressed pair, which is acted upon by an
effective hamiltonian $\tilde{H}$. The change from the full many-body H to $%
\tilde{H}$ is the canonical transformation we were looking for. However, the
scattering operator $S$ is of the form $S=W_{eff}+F,$ where $W_{eff}$ is the
effective interaction between dressed holes, while $F$ is a forward
scattering operator, diagonal in the pair indices $m$ ,$m^{\prime }$ which
accounts for the self-energy corrections of the one-body propagators: it is
evident from (21) that it just redefines the dispersion law $E_{m}$, and,
essentially, renormalizes the chemical potential. It is important to realize
that therefore $F$ must be dropped. This happens also in the Cooper theory%
\cite{kn:kittel}, where an effective interaction involving phonons is
introduced via an (approximate) canonical transformation; the off-diagonal
terms are kept, while the diagonal ones, representing electron self-energy
corrections, are dropped. Therefore the effective Schr\"{o}dinger equation
for the pair reads

\begin{equation}
\left( H_{0}+V+W_{eff}\right) |a>=E_{0}|a>  \label{23}
\end{equation}

and we are interested in the possibility that $E_{0}=2E_{F}-\Delta $, with a
positive binding energy $\Delta $ of the pair. The $V$ interaction just adds
to $W_{eff}$, and this feature allows to include in our model the effects of
other pairing mechanisms, like off-site interactions, inter-planar coupling
and phonons.

We emphasized the fact that the canonical transformation is exact because in
this way our argument does not require U/t to be small. In practice,
however, what we can do is to calculate $W_{eff}$ neglecting 6-body and
higher excitations, and keeping in mind that at least the structure of the
solution is exact when expressed in terms of renormalized matrix elements.
So, in the following, we calculate the bare quantities. The $\alpha $ states
are 3 hole-1 electron determinants which carry no quasi-momentum. We write

\begin{equation}
|\alpha >=|\left\| \left( k^{\prime }+q+k_{2}\right) _{+},\bar{k}%
_{2-},-q_{-},-k_{-}^{\prime }\right\| >  \label{24}
\end{equation}

where $\bar{k}_{2}$ is the electron state and pedices refer to the spin
direction; those with opposite spin indices contribute similarly and yield a
factor of 2 at the end. From the interaction matrix element

\begin{eqnarray}
&<&\left\| \left( k^{\prime }+q+k_{2}\right) _{+},\bar{k}%
_{2-},-q_{-},-k_{-}^{\prime }\right\| |W|d[s]>=  \nonumber \\
&&\delta \left( q-s\right) U\left( q+k^{\prime }+k_{2},-k^{\prime
},s,k_{2}\right)  \nonumber \\
&&-\delta \left( k^{\prime }-s\right) U\left( q+k^{\prime
}+k_{2},-q,s,k_{2}\right)  \label{25}
\end{eqnarray}

we find that the product in the numerator of (22) yields 4 terms; two are
proportional to $\delta (p-s)$ and belong to F, while the cross terms yield
identical contributions to $W_{eff}$. Using (5,6), we obtain the effective
interaction between W=0 pairs:

\[
<\Phi _{\eta }[p]|W_{eff}|\Phi _{\eta }[s]>=4{\sum_{R\in C_{4v}}}\chi
^{\left( \eta \right) }\left( R\right) {\sum_{k_{2}:\varepsilon
(Rs+p+k_{2})>E_{F}}^{o}} 
\]
\[
\frac{U\left( p,k_{2},Rs+p+k_{2},-Rs\right) U\left(
Rs+p+k_{2},-p,Rs,k_{2}\right) }{\varepsilon \left( Rs+p+k_{2}\right)
+\varepsilon \left( s\right) +\varepsilon \left( p\right) -\varepsilon
\left( k_{2}\right) -E_{0}} 
\]

The sum is over occupied $k_{2}$with empty $Rs+p+k_{2}$. Note that $W_{eff}$
does not depend on the sign of $U$. The diagonal elements recover the $%
\Delta $ expression derived from perturbation theory for clusters\cite
{kn:cb5} if $E_{0}$ is replaced by the unperturbed eigenvalue $2\varepsilon
\left( p\right) $.

Since actually $V=0$ in our model, we drop it in the following. The $m$ and $%
m^{\prime }$ indices run over the projected eigenstates $\Phi _{\eta }[k]$
of the kinetic energy, and the k labels run over 1/8 of the BZ. We denote
such a set of empty states $e/8$, and cast the result in the form of a
(Cooper-like) Schr\"{o}dinger equation

\begin{equation}
2\varepsilon \left( k\right) a\left( k\right) +\stackrel{e/8}{
\sum_{k^{\prime }}}W_{eff}\left( k,k^{\prime }\right) a\left( k^{\prime
}\right) =E_{0}a\left( k\right)  \label{31}
\end{equation}

for a self-consistent calculation of $E_{0}$ (since $W_{eff}$ depends on the
solution). Let $N_{C}$ be the number of cells in the crystal. The $U$ matrix
elements scale as $N_{C}^{-1}$ and therefore $W_{eff}$ scales in the same
way. For an infinite system, $N_{C}\rightarrow \infty $ , this is a well
defined, but quite intractable integral equation. The problem becomes
discrete when working with a supercell of $N_{SC}\times N_{SC}=N_{C}$ cells,
with periodic boundary conditions. However, $\Delta $ depends on $U^{\prime
}s$ and $N_{C}$ in a complicated way, within the range of attainable
supercell sizes. We define the Uniform Interaction Model (UIM) in which a
constant effective interaction $V$ prevails for $k$ and $k^{\prime }$ in 1/8
of the empty part of the BZ. Having computed $\Delta $ for a given filling
and $N_{SC}$ according to (26) we can determine the value of $V$ which gives
the same $\Delta $ in the UIM. This will be $V_{eff}$, that is, the
effective $V$ of our theory, characterizing the strength of the attraction
by a single quantity. Since $V_{eff}$ is not strongly size dependent, its
numerical convergence with increasing $N_{SC}$ is observed. In addition, the
UIM can be solved in the thermodynamic limit; writing $E_{0}=2E_{F}-\Delta $
we can estimate $\Delta _{asympt}$ for $N_{C}\rightarrow \infty $ . Indeed,
the Cooper-like Schr\"{o}dinger equation (26) with $W_{eff}=-\frac{V}{N_{C}}$%
, $V>0$, leads to

\begin{equation}
\frac{8}{V}=\int^{0}_{E_{F}}\frac{d\epsilon \rho \left( \varepsilon \right) 
}{2\left( \varepsilon -E_{F}\right) +\Delta _{asympt}}  \label{32}
\end{equation}

where $\rho $ is the density of states, which is solved numerically. We use
as input data the current estimates (in eV) t=1.3, $\varepsilon _{p}$=3.5, $%
\varepsilon _{d}$=0, $U_{p}=6s$, $U_{d}=5.3s$, where s is a scale factor
induced by renormalization.

At s=2.121, with $E_{F}$=-1.3 eV, we get for $^{1}B_{2}$ pairs the results
shown in Table II. Here, $n_{tot}$ is the filling and $V_{eff}$ is derived
by comparison with UIM calculations. We see that $V_{eff}$ is fairly stable
when increasing the supercell size and corresponds to $\Delta _{asympt}$
values of about 20 meV. Convergence to the thermodynamic limit is achieved,
and implies a Cooper-like instability of the Fermi liquid in this model of
the Cu-O plane. $\Delta _{asympt}$ values in the range of several tens to a
few hundreds of meV are obtained with a scale s which is somewhat larger
than 1; it could be that the current estimates of $U^{\prime }$s are a bit
low; however, since the screening excitations are explicitly accounted for
in the Hamiltonian, it is reasonable that the input $U^{\prime }$s must be
somewhat larger than the fully screened interaction. Moreover, contributions
from phonons and other mechanisms can be included by a non zero $V$, and
must be relevant for a comparison with experiment. We find that $^{1}A_{2}$
pairs are more tightly bound close to half filling, but $^{1}B_{2}$ pairs
are favored when the filling increases. We get attraction and pairing at all
fillings: since the present mechanism is driven by symmetry it works
unless the system distorts. Further data will be presented elsewhere.

This work has been supported by the Is\-ti\-tu\-to Na\-zio\-na\-le di Fisica
della Materia. We gratefully acknowledge A. Sa\-gnot\-ti, Universit\`{a} di
Roma Tor Ver\-ga\-ta, for useful and stimulating discussions.

%\newpage

%\newpage

\smallskip

{\bf Tables}

\begin{table}[tbp]
\begin{center}
\begin{tabular}{lclclclclclcl}
C$_{4v}$ & E & C$_{2}$ & 2C$_{4}$ & 2$\sigma $ & 2$\sigma ^{\prime }$ &  & 
&  &  &  &  &  \\ 
A$_{1}$ & 1 & 1 & 1 & 1 & 1 &  &  &  &  &  &  &  \\ 
A$_{2}$ & 1 & 1 & 1 & -1 & -1 & $R_{z}$ &  &  &  &  &  &  \\ 
B$_{1}$ & 1 & 1 & -1 & 1 & -1 & $x^{2}-y^{2}$ &  &  &  &  &  &  \\ 
B$_{2}$ & 1 & 1 & -1 & -1 & 1 & $xy$ &  &  &  &  &  &  \\ 
E & 2 & -2 & 0 & 0 & 0 & $\left( x,y\right) $ &  &  &  &  &  & 
\end{tabular}
\end{center}
\caption{The Character Table of the $C_{4v}$ Group}
\label{Table I}
\end{table}

\narrowtext
\begin{table}[tbp]
\begin{center}
\begin{tabular}{lclclclcl}
$N_{SC}$ & $n_{tot}$ & $\Delta \left( meV\right) $ & $V_{eff}\left(
eV\right) $ & $\Delta _{asympt}\left( meV\right) $ &  &  &  &  \\ 
18 & 1.13 & 121.9 & 7.8 & 41.6 &  &  &  &  \\ 
20 & 1.16 & 42.2 & 5. & 9.0 &  &  &  &  \\ 
24 & 1.14 & 59.7 & 7. & 28.9 &  &  &  &  \\ 
30 & 1.14 & 56. & 5.7 & 13.2 &  &  &  &  \\ 
40 & 1.16 & 30.5 & 6.6 & 23.4 &  &  &  & 
\end{tabular}
\end{center}
\caption{Binding Energy of W=0 Pairs.}
\label{Numerical Results.}
\end{table}
\bigskip

\end{document}